\documentclass[epj,referee]{svjour}
\usepackage{graphics}
\begin{document}
\title{Type-II intermittency characteristics in the presence of noise}
\author{Alexey A.~Koronovskii\inst{1} \and Alexander E.~Hramov\inst{1}% etc
%\thanks{\emph{Faculty of Nonlinear Processes, Saratov State
%University - Astrakhanskaya, 83, Saratov, 410012, Russia}}%
}                     % Do not remove
\institute{Faculty of Nonlinear Processes, Saratov State University
- Astrakhanskaya, 83, Saratov, 410012, Russia, \\ e-mail: alkor@nonlin.sgu.ru}% \and the second here}
\date{Received: March 26, 2008 / Revised version: date}
\abstract{ We consider a type of intermittent behavior that occurs
as the result of the interplay between dynamical mechanisms giving
rise to type-II intermittency and random dynamics. We analytically
deduce the law for the distribution of the laminar phases, which has
never been obtained hitherto. The already known  dependence of the
mean length of the laminar phases on the criticality parameter [PRE
\textbf{68} (2003) 036203] follows as a corollary of the carried out
research. We also prove that this dependence obtained earlier under
the assumption of the fixed form of the reinjection probability does
not depend on the relaminarization properties, and, correspondingly,
the obtained expression of the mean length of the laminar phases on
the criticality parameter remains
correct for different types of the reinjection probability.%
 % end of PACS codes
} %end of abstract
\maketitle
\section{Introduction}
\label{sct:intro}

Intermittency is known to be an ubiquitous
phenomenon, with its arousal and main statistical properties having
been studied and characterized already since long time ago. The
different types of intermittency have been classified as types
I--III~\cite{Dubois:1983_IntermittencyIII}, on--off
intermittency~\cite{Platt:1993_intermittency,Heagy:1994_intermittency,%
Hramov:2005_IGS_EuroPhysicsLetters}, eyelet
intermittency~\cite{Pikovsky:1997_EyeletIntermitt,%
Boccaletti:2002_LaserPSTransition_PRL} and ring
intermittency~\cite{Hramov:RingIntermittency_PRL_2006,Hramov:2007_2TypesPSDestruction}.
From the other side, increasing interest has been put recently in
the study of the constructive role of noise and fluctuations in
nonlinear systems. In particular, it was discovered that random
fluctuations can actually induce some degree of order
in a large variety of nonlinear systems~\cite{Pikovsky:1997_CoherenceResonance,%
Mangioni:1997_Noise,Hramov:2006_PLA_NIS_GS}, and such  phenomena
were widely observed in relevant physical
circumstances~\cite{Boccaletti:2002_LaserPSTransition_PRL,Zhou:PRE2003}.

There are no doubts that different types of intermittent behavior
may take place in the presence of noise and fluctuations in a wide
spectrum of systems, including cases of practical interest for
applications in physical, radio engineering and other applied
sciences. It is plausible that such an interaction would originate
new types of dynamics. Therefore, the intermittent behavior in the
presence of noise has been studied by means of Fokker-Plank
equation~\cite{Hirsch:1982_Intermittency} and adopting
renormalization group analysis~\cite{Hirsch:1982_IntermittencyPLA},
but the characteristic relations were obtained only in the
subcritical region, where the intermittent behavior is observed both
in the presence of noise and without noise.
Recently~\cite{Kye:2000_TypeIAndNoise,Cho:2002_TypeINoseExpeiment,Hramov:2007_TypeIAndNoise},
the theoretical and experimental consideration of the intermittent
behavior in the presence of noise has been considered in the
supercritical region (where intermittency does not take place in the
absence of noise) for the type-I intermittency.

Obviously, in the presence of noise the other types of intermittent
behavior mentioned above may result in the distinct types of
dynamics. In this paper we report for the first time the important
characteristic (namely, the distribution of the laminar phase
lengths) deduced analytically for the type-II intermittency in the
presence of noise for the region of the supercritical parameter
values, which has never been obtained hitherto. This characteristic
is of great importance, since scientists studying the intermittent
behavior of the complex systems often can not variate the
criticality parameter in the experiments (e.g., in physiological and
biological systems), and, therefore, they can deal only with the the
distribution of the laminar phase lengths, whereas the dependence of
the mean length of the laminar phases on the criticality parameter
can not be obtained. The already known dependence of the mean length
of the laminar phases on the criticality
parameter~\cite{Kye:2003_TypeIIAndIIINoiseExperiment} follows as a
corollary of the carried out research. Moreover, we prove that this
dependence obtained in~\cite{Kye:2003_TypeIIAndIIINoiseExperiment}
under the assumption of the fixed reinjection probability (taken in
the forms of delta-function and uniform distribution) does not
depend practically on the relaminarization properties, and,
correspondingly, the obtained expression of the mean length of the
laminar phases on the criticality parameter remains correct for
different forms of the reinjection probability. The obtained
analytical distribution of the laminar phase length is verified by
means of numerical calculations of the model system dynamics.

\section{Analytical approach}
\label{sec:AnalyticalApproach}

The standard model that is used to study the type-II
intermittency~\cite{Kye:2003_TypeIIAndIIINoiseExperiment} is the
one-parameter cubic map
\begin{equation}
x_{n+1}=f(x_n)=(1+\epsilon)x_n+x_n^3, \label{eq:CubicMap}
\end{equation}
where $\epsilon$ is a control parameter. Below the critical
parameter value (i.e., for $\epsilon<\epsilon_c$), the stable fixed
point $x_{c}=0$ is observed, while above $\epsilon_c$ this fixed
point $x_c=0$ becomes unstable and the point representing the state
of the map~(\ref{eq:CubicMap}) moves around it with slowly
increasing amplitude. This movement in the vicinity of the fixed
point corresponds to the laminar phase, its mean length $T$ being
inversely proportional to $(\epsilon-\epsilon_c)$, i.e.
\begin{equation}\label{eq:Type-IIntermittencyPowerLaw}
T\sim(\epsilon-\epsilon_c)^{-1}.
\end{equation}

To develop the theory of type-II intermittency in the presence of
noise, we consider the same cubic map~(\ref{eq:CubicMap}) with the
addition of a stochastic term $\xi_n$
\begin{equation}
x_{n+1}=f(x_n)=(1+\epsilon)x_n+x_n^3+\xi_n,
\label{eq:StochasticCubicMap}
\end{equation}
where $\xi_n$ is supposed to be a delta-correlated white noise
[${\langle\xi_n\rangle=0}$,
${\langle\xi_n\xi_m\rangle=D\delta(n-m)}$].

The influence of the stochastic term $\xi_n$ on the behavior of the
system is governed by the value of parameter $D$. For positive
values of the control parameter $\epsilon$ ($\epsilon>0$), the point
corresponding to the behavior of
system~(\ref{eq:StochasticCubicMap}) moves in the iteration diagram
around the unstable fixed point, its motion being perturbed by the
stochastic force. As far as the intensity of the noise is not large,
the characteristics being close to the classical type-II
intermittency are observed.

A different scenario occurs for control parameters $\epsilon$
assuming negative values ($\epsilon=-\varepsilon$, where
$\varepsilon=|\epsilon|>0$). In this case, the point corresponding
to the behavior of system~(\ref{eq:StochasticCubicMap}) is localized
for a long time in the region
${-\sqrt{\varepsilon}<x<\sqrt{\varepsilon}}$ and its dynamics is
also perturbed by the stochastic force. As soon as the system state
point arrives at one of the boundaries $x_b=\pm\varepsilon^{-1/2}$
due to the influence of noise, a turbulent phase arises, though such
kind of events is very rare.

In this case, the behavior of the map~(\ref{eq:StochasticCubicMap})
differs radically from the dynamics of the
system~(\ref{eq:CubicMap}), since the turbulent phases are not
observed for $\epsilon<0$ if there is no noise. Therefore, such a
region of negative values of the $\epsilon$-parameter is the main
subject of interest for the type-II intermittency in the presence of
noise.

\subsection{Probability density}
\label{sbsec:ProbabilityDensity}

Having supposed that: (i) the value of $\epsilon$ is negative and
rather small and (ii) the value of $x$ changes per one iteration
insufficiently, we can consider ${(x_{n+1}-x_{n})}$ as the time
derivative $\dot x$ and undergo from the system with discrete
time~(\ref{eq:StochasticCubicMap}) to the flow system, in the same
way as in the case of the classical theory of the
intermittency~\cite{Dubois:1983_IntermittencyIII}.

Since the stochastic term is present
in~(\ref{eq:StochasticCubicMap}) we have to examine the stochastic
differential equation
\begin{equation}
dX=X(X^2-\varepsilon)\,dt+dW \label{eq:SDEType2}
\end{equation}
(where $X(t)$ is a stochastic process, $W(t)$ is a one-dimensional
Winner process, {$\varepsilon=|\epsilon|$}) instead of the ordinary
differential equation ${dx/dt=x(x^2+\epsilon)}$ considered in the
classical theory of type~II intermittency.

The stochastic differential equation~(\ref{eq:SDEType2}) is
equivalent to the Fokker-Plank equation
\begin{equation}
\frac{\partial \rho_X}{\partial t}=-\frac{\partial}{\partial
x}\left(x(x^2-\varepsilon)\rho_X\right)+\frac{D}{2}\frac{\partial^2\rho_X}{\partial
x^2} \label{eq:FPEMinusType2}
\end{equation}
for the probability density $\rho_X(x,t)$ of the stochastic process
$X(t)$. The chosen initial condition is
${\rho_X(x,0)=\delta(x-\Delta)}$, where $\delta(\cdot)$ is a
delta-function. Such a choice of the initial form of the probability
density $\rho_X(x,0)$ corresponds to the beginning of the laminar
phase, when the point representing the state of the
system~(\ref{eq:StochasticCubicMap}) is in the place with coordinate
$x=\Delta$ ($-\varepsilon^{1/2}<\Delta<\varepsilon^{1/2}$) at time
$t=0$. In other words, we suppose that the reinjection probability
is a $\delta$-function
\begin{equation}
P_{in}(x)=\delta(x-\Delta)
\end{equation}
and after the relaminarization process the system is always returned
to the state $x=\Delta$. Although the reinjection probability
$P_{in}(x)$ is well-known to be important factor and should be taken
into account when the statistical properties of the intermittent
behavior are
studied~\cite{Kin:1994_NewIntermittencyCharacteristics,%
Kim:1998_IntermittencyCharacteristicsPRL}, in the considered problem
the form of the reinjection probability practically does not
influence on the distribution of the laminar phase lengths (and the
dependence of the mean laminar phase length on the criticality
parameter, respectively), as it will be shown below.

To reduce the number of the control parameters the normalization
$z=x/\sqrt{\varepsilon}$, $\tau=t\varepsilon$ may be used, after
which Eq.~(\ref{eq:FPEMinusType2}) may be rewritten in the form
\begin{equation}
\frac{\partial \rho_Z}{\partial \tau}=-\frac{\partial}{\partial
z}\left(z(z^2-1)\rho_Z\right)+\frac{D^*}{2}\frac{\partial^2\rho_Z}{\partial
z^2}, \label{eq:NormFPEMinusType2}
\end{equation}
where $D^*=D\varepsilon^{-2}$,
$\rho_Z(z,\tau)=\rho_X(z{\varepsilon}^{1/2},\tau{\varepsilon}^{-1})$.

As the coordinate of the system state stays for a long time in the
region $|z|<z_b=1$, one can  suppose that the probability density
may found the form of the metastable distribution decaying slowly
for a long period of time. The relaxation process of the probability
density to this metastable state is supposed to be very fast in
comparison with the time of the metastable distribution decay,
therefore, one can neglect the transient ${0\leq t\le t_{tr}}$.
Under the assumptions made above the probability density may be
written in the form ${\rho_Z(z,\tau)=A(\tau)g(z)}$, ${\forall
z\in(-1;1)}$, where $A(\tau)>0$ decreases very slowly as time
increases, i.e. $dA/d\tau\approx 0$. The function $g(z)$ should
satisfy the conditions
\begin{equation}\label{eq:ConditionsForPType2}
g(z)>0 ~~ \forall z\in(-1;1) \quad\mathrm{and}\quad
\displaystyle\int\limits_{-1}^{1} g(z)\,dz<\infty.
\end{equation}
As the maximum of the probability density should coincide with the
stable fixed point $z_s=0$, one has
\begin{equation}\label{eq:ExtremumConditionType2}
g'(z)\left|_{z=0}\right.=0.
\end{equation}

Under the mentioned assumption, we consider the ordinary
differential equation
\begin{equation}
D^*g''(z)-2\left(z(z^2-
1)g(z)\right)'=0\label{eq:ODUSecondOrderType2}
\end{equation}
instead of~(\ref{eq:NormFPEMinusType2}) for the region $z\in(-1;1)$.

This equation is equivalent to
\begin{equation}
D^*g'(z)-2z(z^2- 1)g(z)+C_1=0, \label{eq:ODUFirstOrderType2}
\end{equation}
where $C_1$ is constant. To solve this equation we use the
integrating factor
\begin{equation}
M(z)=\exp\left(\frac{z^2}{D^*}\left(1-\frac{z^2}{2}\right)\right).
\end{equation}
The solution of~(\ref{eq:ODUFirstOrderType2}) may be found in the
form
\begin{equation}\label{eq:SolutionOfODEType2}
g(z)=\frac{\displaystyle
2C_1\int_0^z\exp\left(\frac{s^2}{D^*}\left(1-\frac{s^2}{2}\right)\right)\,ds+C_2}{\displaystyle
D^*\exp\left(\frac{z^2}{D^*}\left(1-\frac{z^2}{2}\right)\right)}.
\end{equation}
From Eq.~(\ref{eq:SolutionOfODEType2}) one can obtain easily that
${g'(0)=2C_1/D^*}$. Taking into account the
condition~(\ref{eq:ExtremumConditionType2}), one comes to the
conclusion that $C_1\equiv 0$. Note, in this case the obtained
function
\begin{equation}\label{eq:GFunctionType2}
g(z)=\frac{C_2}{D^*}\exp\left(-\frac{z^2}{D^*}\left(1-\frac{z^2}{2}\right)\right).
\end{equation}
also satisfies the conditions~(\ref{eq:ConditionsForPType2}).
Therefore, the probability density $\rho_Z(z,\tau)$ in the region
$z\in(-1;1)$ is
\begin{equation}
\rho_Z(z,\tau)\simeq
A(\tau)\exp\left(-\frac{z^2}{D^*}\left(1-\frac{z^2}{2}\right)\right).
\end{equation}

When the laminar phase is interrupted, the system escapes from a
metastable state. Therefore, we suppose that the decrease of
$A(\tau)$ should be determined by the probability distribution taken
in the boundary points $z=\pm 1$, i.e., $dA(\tau)/d\tau\sim
-\rho_Z(\pm 1,\tau)$. This assumption, which is also equivalent to
neglecting the time correlation of the orbit, may be rewritten as
\begin{equation}
\frac{dA(\tau)}{d\tau}=-kA(\tau)\exp\left(-\frac{1}{2D^*}\right),
\end{equation}
where $k$ is a proportionality coefficient. Evidently, the decrease
of $A(\tau)$ is described by the exponential law
\begin{equation}
A(\tau)=A(0)\exp(-k\eta\tau),\quad \eta=\exp{(-1/(2D^*))}.
\end{equation}

Having returned to the initial variables $x$ and $t$ we derive the
following expression for the probability density
\begin{equation}\label{eq:ProbabDistribMinusType2}
\rho_X(x,t)\simeq A(t)
\exp\left(-\frac{x^2}{D}\left(\varepsilon-\frac{x^2}{2}\right)\right),
\end{equation}
where
\begin{equation}\label{eq:AvsTType2}
A(t)=A(0)\exp\left(-\frac{t}{T}\right),
\end{equation}
and
\begin{equation}\label{eq:TheoreticalLawForMeanLenghtType2}
T=\frac{1}{k\varepsilon}\exp\left(\frac{\varepsilon^{2}}{2D}\right),
\end{equation}
with $A(t)$ being considered as a normalizing factor, i.e.,
\begin{equation}\label{eq:NormalizationConditionForAType2}
A(0)\int\limits_{-\sqrt{\varepsilon}}^{\sqrt{\varepsilon}}
\exp\left(-\frac{x^2}{D}\left(\varepsilon-\frac{x^2}{2}\right)\right)\,dx=1.
\end{equation}

To confirm the assumptions made above and the obtained equations, we
have compared the evolution of the probability density $\rho_X(x,t)$
given by~(\ref{eq:ProbabDistribMinusType2}) with the result of the
direct numerical calculation of the Fokker-Plank
equation~(\ref{eq:FPEMinusType2}) with the values of control
parameters $\varepsilon=10^{-2}$, $D=10^{-5}$.

\begin{figure}
\centerline{\resizebox{0.35\textwidth}{!}{%
  \includegraphics{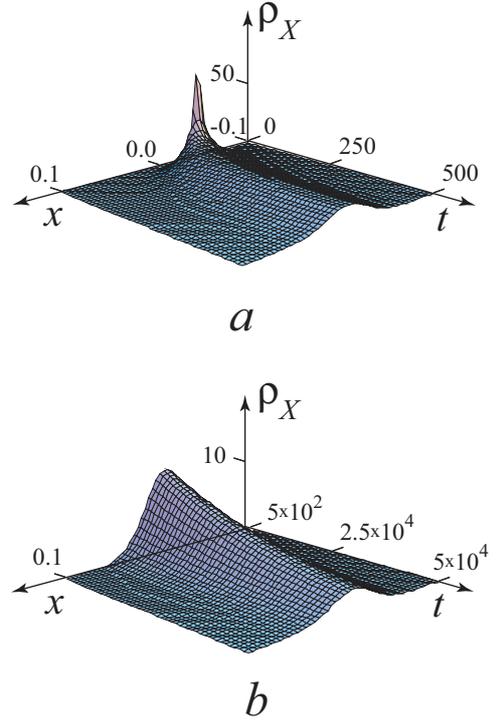}}
} \caption{The evolution of the probability density $\rho_X(x,t)$
obtained by means of the direct numerical integration of
Fokker-Plank equation~(\ref{eq:FPEMinusType2}),
${\varepsilon=10^{-2}}$, ${D=10^{-5}}$. (\textit{a})~The initial
fragment of the density evolution involving the transient (${0\leq
t<t_{tr}}$, $t_{tr}\approx 3\times10^{2}$). (\textit{b})~The
long-time evolution of $\rho_X(x,t)$, with the transient being
omitted, $5\times10^{2}\leq t\leq 5\times10^{4}$}
\label{fgr:ProbDensityMinusEvolutionType2}
\end{figure}

The evolution of the probability density $\rho_X(x,t)$ obtained by
the numerical calculation of~(\ref{eq:FPEMinusType2}) is shown in
Fig.~\ref{fgr:ProbDensityMinusEvolutionType2}. One can see that
after the very short transient ${0\leq t\le t_{tr}}$ the probability
density $\rho_X(x,t)$ arrives the state being close to stationary
(Fig.~\ref{fgr:ProbDensityMinusEvolutionType2}\,\textit{a}). After
that the value of $\rho_X(x,t)$ decreases very slowly (according to
the exponential law) with time increasing, with the form of the
dependence of the probability density on $x$-coordinate being
invariable
(Fig.~\ref{fgr:ProbDensityMinusEvolutionType2}\,\textit{b}).

\begin{figure}
\centerline{\resizebox{0.35\textwidth}{!}{%
  \includegraphics{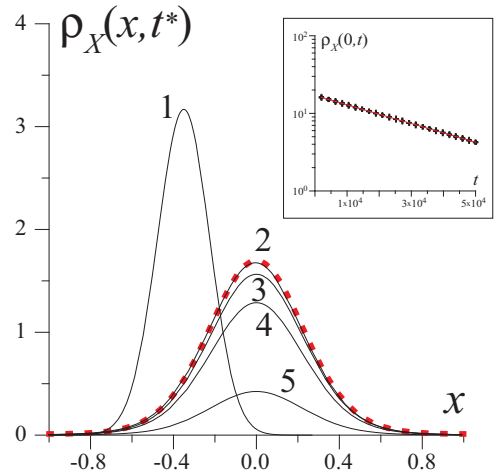}}
} \caption{The profiles of the probability density $\rho_X(x,t^*)$
taken in the different moments of time $t^*$ obtained from the
direct numerical calculation of Fokker-Plank
equation~(\ref{eq:FPEMinusType2}). In the frame the dependence of
the probability density $\rho(0,t)$ in the stationary point $x=0$ on
time $t$ ($+$) and its exponential approximation (solid line) are
shown in the logarithmic scale}
\label{fgr:ProbDensityMinusProfilesType2}
\end{figure}

Fig.~\ref{fgr:ProbDensityMinusProfilesType2} also shows the profiles
of the probability density $\rho_X(x,t^*)$ taken in the different
moments of time. It is evident, that after a very short transient
(curve~1, $t_1^*=15$), the density $\rho_X(x,t)$ practically does
not change when time increases.

Two different profiles $\rho_X(x,t^*)$ corresponding to the time
moments $t^*_2=5\times10^2$ and $t^*_3=3\times10^3$ (curves~2 and 3,
respectively) are very close to each other despite of the large time
interval $\Delta t={t^*_3-t^*_2}$ between them. Moreover, they are
in  very good agreement with the approximated solution $A(0)g(x)$
described by Eq.~(\ref{eq:ProbabDistribMinusType2}) and shown in
Fig.~\ref{fgr:ProbDensityMinusProfilesType2} by means of squares. As
time goes on, the amplitude of the probability density decreases
according to the exponential law, but very slowly (see
Fig.~\ref{fgr:ProbDensityMinusProfilesType2}, curves~4 and 5,
$t^*_4=10^4$ and $t^*_5=5\times10^4$, respectively), although the
probability density form remains the same for all  times.

Therefore, taking into account the results of the direct numerical
calculations of Fokker-Plank equation~(\ref{eq:FPEMinusType2}) and
the comparison with the obtained approximated
solution~(\ref{eq:ProbabDistribMinusType2}), we come to  the
conclusion that our assumptions  are correct and can be used for the
further analysis.

The evolution of the probability density $\rho_X(x,t)$ may be
considered separately on two time intervals ${0\leq t < t_{tr}}$ and
${t_{tr}\leq t < +\infty}$, respectively. The first time interval
corresponds to the transient when the probability density
$\rho_X(x,t)$ evolves to the form~(\ref{eq:ProbabDistribMinusType2})
being close to stationary. Only when ${0\leq t < t_{tr}}$ the form
of the reinjection probability $P_{in}(x)$ may influence on the
evolution of the probability density $\rho_X(x,t)$. For  $t\geq
t_{tr}$ (when the transient is elapsed), the evolution of the
probability density is defined completely by
Eq.~(\ref{eq:ProbabDistribMinusType2}) and it does not depend
entirely on the reinjection probability $P_{in}(x)$. Since the
transient is very short in comparison with the exponential decrease
of the probability density $\rho_X(x,t)$ we can neglect them and use
only the second time interval ${t_{tr}\leq t < +\infty}$ to obtain
the statistical characteristics of the type-II intermittent behavior
in the presence of noise. It is clear, that in this case the
obtained results do not depend on the relaminarization process and
the reinjection probability $P_{in}(x)$.

\subsection{Distribution of the
laminar phase lenghts} \label{sbsec:DistributionPhaseLenghts}

Let us discuss now the relationship between the probability density
$\rho_X(x,t)$ and the distribution $p(t)$ of the laminar phase
lenghts $t$. For this purpose we consider the ensemble of
systems~(\ref{eq:SDEType2}). Let there be $N(t)$ systems which at
the moment of time $t$ are in the stage of the laminar phase.
Obvioulsy, the states $x$ of these systems should be distributed
over interval $(-\sqrt{\varepsilon},\sqrt{\varepsilon})$ according
to the probability density $\rho_X(x,t)$. After the infinitely small
time interval $dt$ several systems stop demonstrating the laminar
behavior (the turbulent phase begins in these systems) and,
correspondingly, their states $x$ leave the interval
$(-\sqrt{\varepsilon},\sqrt{\varepsilon})$, while the states of
other $N(t+dt)$ systems demonstrating the laminar dynamics as before
are distributed in accordance with the probability density
$\rho_X(x,t+dt)$). If the total number of the systems in the
ensemble under consideration is $N_0$, the number of the systems in
which the laminar phase is finished during time interval $[t,t+dt)$
is
\begin{equation}
\begin{array}{l}
\displaystyle N(t)-N(t+dt)=\\
\displaystyle\qquad
=N_0\int\limits_{-\sqrt{\varepsilon}}^{\sqrt{\varepsilon}}[\rho_X(x,t)-\rho_X(x,t+dt)]\,dx.
\end{array}
\end{equation}
Since the number of the systems in the ensemble in which the laminar
phase is interrupted at the moment of time falling in the time
interval between $t$ and $t+dt$ is connected with the distribution
$p(t)$ of the laminar phase lengths $t$ as
\begin{equation}
N(t)-N(t+dt)=N_0p(t)\,dt,
\end{equation}
one can obtain the relationship
\begin{equation}
p(t)=-\int\limits_{-\sqrt{\varepsilon}}^{\sqrt{\varepsilon}}\frac{\partial\rho_X(x,t)}{\partial
t}\,dx
\end{equation}
between the probability density $\rho_X(x,t)$ and the distribution
$p(t)$ of the laminar phase lenghts.

Using relations~(\ref{eq:ProbabDistribMinusType2}),
(\ref{eq:AvsTType2}) and (\ref{eq:NormalizationConditionForAType2})
one can obtain, that the laminar phase distribution is governed by
the exponential law
\begin{equation}\label{eq:LamPahseLengthDistributionType2}
p(t)=T^{-1}\exp\left(-{t/T}\right),
\end{equation}
where $T$ defined by Eq.~(\ref{eq:TheoreticalLawForMeanLenghtType2})
is the mean length of the laminar phases. The obtained
expression~(\ref{eq:TheoreticalLawForMeanLenghtType2}) for the mean
length $T$ of the laminar phases is consistent with the formal
solution derived in the previous
considerations~\cite{Kye:2003_TypeIIAndIIINoiseExperiment,%
Pikovsky:1983_IntermittencyWithNoise}, that may be considered as an additional evidence of the correctness of our results. Nevertheless, we would like to emphasize that the applicability of the results obtained earlier analytically \cite{Kye:2003_TypeIIAndIIINoiseExperiment,%
Pikovsky:1983_IntermittencyWithNoise} are limited greatly by the
assumptions (made to deduce corresponding equation) concerning the
character of the reinjection process. Indeed, it is well known that
the characteristics of the intermittent behavior even in the case
without noise depend not only on the structure of the local Poincare
map but on the reinjection probability distribution, with this
dependence being sufficient (see e.g.,
\cite{Kin:1994_NewIntermittencyCharacteristics,Kim:1998_IntermittencyCharacteristicsPRL}).
At the same time, the analytical results mentioned above were
obtained only under assumptions of uniform and delta-function
reinjection probability distribution. Our study allows to extend
known relation to all types of the reinjection probability
distribution, since based on the consideration carried out above we
state that Eqs.~(\ref{eq:TheoreticalLawForMeanLenghtType2}) and
(\ref{eq:LamPahseLengthDistributionType2}) do not depend on the
relaminarization process properties and may be used for the
arbitrary reinjection probability $P_{in}(x)$.

\section{Numerical results}
\label{sec:NumericalResults}

To verify the obtained theoretical predictions, we consider
numerically the intermittent behavior of the quadratic
map~(\ref{eq:StochasticCubicMap}) with a stochastic force. The
reinjection procedure has been fulfilled as follows: when the value
of the $x_n$-variable leaves the interval ${-1<x<+1}$, the next its
value has been taken as $x_{n+1}=10^{-2}x_n$. Since the dependence
of the mean laminar phase length on the criticality
parameter~(\ref{eq:TheoreticalLawForMeanLenghtType2}) has already
been studied~\cite{Kye:2003_TypeIIAndIIINoiseExperiment}, in our
calculations we focus on the consideration of distribution of the
laminar phase lengths.

Obviously, if the intensity of noise $D$ is equal to zero the
intermittent behavior is observed for $\epsilon>0$, whereas the
stable fixed point takes place for $\epsilon<0$. Having added the
stochastic force we can expect that the intermittent behavior may be
also observed in the area of the negative values of the criticality
parameter $\epsilon$.

\begin{figure}
\centerline{\resizebox{0.35\textwidth}{!}{%
  \includegraphics{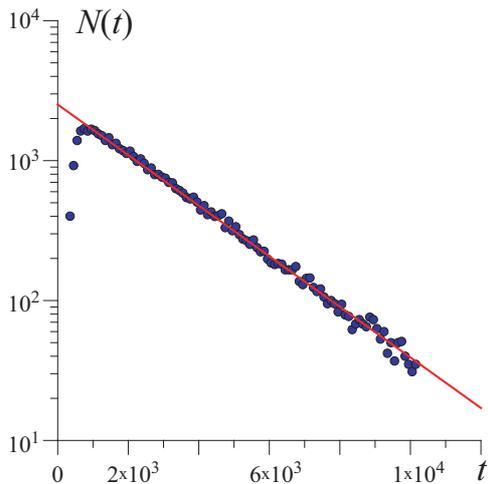}}
} \caption{The distribution of the laminar phase lengths for
map~(\ref{eq:StochasticCubicMap}). The criticality parameter value
has been selected as $\epsilon=-10^{-3}$, the noise intensity is
$D=10^{-6}$. The ordinate axis is presented in the logarithmic
scale. The theoretical exponential
law~(\ref{eq:LamPahseLengthDistributionType2}) is shown by the solid
line} \label{fgr:TestQuadraticMapMeanLengthMinus}
\end{figure}

The distribution of the laminar phase lengths  is in very good
accordance with the exponential
law~(\ref{eq:LamPahseLengthDistributionType2}) predicted by the
theory of the type-II intermittency with noise (see
Fig.~\ref{fgr:TestQuadraticMapMeanLengthMinus}). Note the presence
of the small region of the short laminar phase lengths in
Fig.~\ref{fgr:TestQuadraticMapMeanLengthMinus} where the deviation
from the prescribed exponential
law~(\ref{eq:LamPahseLengthDistributionType2}) is observed. This
region corresponds to the transient ${0\leq t < t_{tr}}$ when the
probability density $\rho_X(x,t)$ evolves to the
form~(\ref{eq:ProbabDistribMinusType2}) being close to stationary as
it was discussed above. The existence of this transient time
interval does not influence practically on the
characteristics~(\ref{eq:TheoreticalLawForMeanLenghtType2}) and
(\ref{eq:LamPahseLengthDistributionType2}) of the intermittent
behavior in the presence of noise in the full agreement with the
conclusions made above. So, the intermittent behavior observed in
the quadratic map with the stochastic force agrees well with the
theoretical predictions.

\section{Conclusion}
\label{sec:Conclusion}

In conclusion, we have reported  a  type of intermittency behavior
caused by the cooperation between the deterministic mechanisms and
random dynamics. The distribution of the laminar phase lengths being
one of the important characteristics of the intermittent dynamics
has been deduced analytically. Having considered the standard model
of type-II intermittency in the presence of noise we can conclude
that (i)~noise induces new features in the intermittent behavior of
a system demonstrating type-II intermittency, with  new dynamical
properties being observed above the former value of the criticality
parameter; (ii)~the results of numerical simulations are in
excellent agreement with the developed theory; (iii)~the
relaminarization process properties and the reinjection probability
do not seem to play a major role for the statistical characteristic
of type-II intermittency, and, the obtained expression of the mean
length of the laminar phases on the criticality parameter as well as
the dependence of the mean length of the laminar phases on the
criticality parameter remain correct for different forms of the
reinjection probability.

Though the characterization of the intermittent process has been
explicitly derived here for model system, we expect that the very
same mechanism can be observed in many other relevant circumstances
where the level of natural noise is sufficient, e.g. in the
physiological~\cite{Hramov:2006_Prosachivanie,Hramov:2006_RAT_ON-OFF,Hramov:2007_UnivariateDataPRE}
or physical systems~\cite{Boccaletti:2002_LaserPSTransition_PRL}.

This work has been supported by U.S.~Civilian Research \&
Development Foundation for the Independent States of the Former
Soviet Union (CRDF, grant {REC--006}), Russian Foundation of Basic
Research (project 07-02-00044), the Supporting program of leading
Russian scientific schools (project NSh--355.2008.2). We thank
``Dynasty'' Foundation. A.E.H. also acknowledges support from the
President Program, Grant No. MD-1884.2007.2.

\end{document}